\documentclass[preprint,aps,amssymb,showpacs,superscriptaddress,nofootinbib]{revtex4}
\usepackage{graphicx}

\newcommand{\Case}[2]{{\textstyle \frac{#1}{#2}}}
\newcommand{\lP}{\ell_{\mathrm P}}

\begin{document}
%
%\today
\preprint{IMSc/2005/01/01}

\title{Discreteness Corrections to the Effective Hamiltonian of Isotropic 
Loop Quantum Cosmology}

\author{Kinjal Banerjee}
\email{kinjal@imsc.res.in}
\affiliation{The Institute of Mathematical Sciences\\
CIT Campus, Chennai-600 113, INDIA.}

\author{Ghanashyam Date}
\email{shyam@imsc.res.in}
\affiliation{The Institute of Mathematical Sciences\\
CIT Campus, Chennai-600 113, INDIA.}

\begin{abstract}

One of the qualitatively distinct and robust implication of Loop Quantum
Gravity (LQG) is the underlying discrete structure. In the cosmological
context elucidated by Loop Quantum Cosmology (LQC), this is manifested
by the Hamiltonian constraint equation being a (partial) difference
equation. One obtains an effective Hamiltonian framework by making the
continuum approximation followed by a WKB approximation.  In the large
volume regime, these lead to the usual classical Einstein equation which
is independent of both the Barbero-Immirzi parameter $\gamma$ as well
as $\hbar$. In this work we present an alternative derivation of the
effective Hamiltonian by-passing the continuum approximation step. As a
result, the effective Hamiltonian is obtained as a close form expression
in $\gamma$. These corrections to the Einstein equation can be thought
of as corrections due to the underlying discrete (spatial) geometry with
$\gamma$ controlling the size of these corrections. These corrections
imply a bound on the rate of change of the volume of the isotropic
universe. In most cases these are perturbative in nature but for
cosmological constant dominated isotropic universe, there are
significant deviations.

\end{abstract}


\maketitle

%%%%%%%%%%%%%%%%%%%%%%%%%%%%%%%%%%%%%%%%%%%%%%%%%%%%%%%%%%%%%%%%%%%%%%%

\section{Introduction}
It is widely believed that one needs to construct a quantum theory of
gravity. In two of the leading approaches, string theory
\cite{StringRef} and LQG \cite{LQGRev}, the marriage of quantum theory
and gravity needs some new essential input(s).  In the ambitious string
theory, one such input is extra dimensions still retaining framework of
continuum space-time geometry. In LQG, the new feature that emerges is
the discrete nature of the spatial {\em geometry} (as distinct from
manifold structure). It is also a fact that for most purposes, classical
continuum geometry and the Einstein equations suffice to an excellent
approximation. It is thus a natural question: How much deviations from
classical general relativity (GR) are predicted from the more refined
quantum versions?

In order to address such a question in some quantitative manner, it is
necessary to be able to identify physical contexts in which the more
refined theory can be computed in sufficiently explicit terms. A natural
context is that of homogeneous cosmologies or even simpler, that of
homogeneous and isotropic cosmologies. In this context the geometry is
described in terms of a single variable of a single coordinate -- the
scale factor -- while matter can be effectively restricted to scalar
field(s) with possibly non-minimal couplings (eg dilaton).  Even within
the cosmological context, there are regimes of small volume (closer to
the classical singularity) and regimes of large volume where classical
GR is expected to be an excellent approximation. In this work, we obtain
deviations from GR in the large volume regime, as implied by Loop
Quantum Cosmology \cite{LQCRev,Bohr}, as a function of the
Barbero-Immirzi parameter $\gamma$ (and another order 1 dimensionless
parameter $\mu_0$).

In the previous works on obtaining an effective Hamiltonian
\cite{SemiClass,FundamentalDisc,EffHamiltonian}, a two step approach was
taken, namely, a continuum approximation wherein the fundamental
difference equation is replaced by a {\em second order} differential
equation (Wheeler-DeWitt equation) followed by a WKB approximation of
the differential equation. We will refer to the first step as the {\em
pre-classical approximation} \cite{FundamentalDisc}. The second order
differential equation is independent of the Barbero-Immirzi parameter
and the Planck constant. One could go beyond the pre-classical
approximation and obtain higher order (in $\gamma$) corrections to the
effective Hamiltonian following the same two step approach. In this
work, however,  we derive an effective Hamiltonian as an explicit closed
form expression involving $\gamma$, by-passing the pre-classical
approximation.

In section \ref{LQCSummarySection}, we recall the relevant aspects of
isotropic LQC namely, the fundamental difference equation, the form of
generic solutions, the pre-classical approximation and the WKB route to
obtaining an effective Hamiltonian constraint. In section
\ref{MainSectionI}, we make the WKB approximation directly at the level
of the fundamental equation to arrive at the effective Hamiltonian as
well as the conditions for validity of the WKB approximation. The
effective Hamiltonian is obtained as a close form expressions in
$\epsilon := 2 \mu_0 \gamma$, where $\mu_0$ is a dimensionless parameter
of order 1.  In section \ref{SolnSection} we discuss some illustrative
solutions of the effective dynamics for minimally coupled single scalar
field. We also discuss, how the solutions of the $\epsilon-$dependent
equations differ if phenomenological matter (dust, radiation and
cosmological constant) are incorporated somewhat naively.  In the limit
$\epsilon \to 0$ these equations and solutions go over to the usual
Einsteinian classical solutions. These solutions exhibit the situations
wherein, the discreteness corrections (codified by $\epsilon$) appear as
perturbative corrections as well as when the discreteness corrections
have qualitatively different implications.  Finally in
\ref{DiscussionSection}, we clarify the nature of approximations and
domains of validity and summarize our results. 

\section{Summary of LQC framework}
\label{LQCSummarySection}

We work within the LQC framework as detailed in \cite{Bohr}.  For our
purposes, it may be summarized as follows. Classically, we have a phase
space with coordinates $c, p$ and the Poisson bracket given by $\{c, p\}
= \Case{\kappa \gamma}{3}$, where $\kappa = 16 \pi G$ and $\gamma$ is
the Barbero-Immirzi parameter whose precise value does not concern us
here. The coordinate $c$ denotes the only invariant part of the
gravitational gauge connection while $p$ denotes the only invariant part
of the densitized triad. The {\em loop} quantization leads to a
kinematical (i.e.  before imposition of the only surviving Hamiltonian
constraint) Hilbert space conveniently described in terms of the
eigenstates of the densitized triad operator ($\lP^2 := \kappa \hbar$),
\begin{equation}\label{TriadAction} 
\hat{p} | \mu \rangle ~ = ~
\Case{1}{6} \gamma \lP^2 \mu | \mu \rangle~~,~~ \langle \mu |
\mu^{\prime} \rangle = \delta_{\mu \mu^{\prime}} 
\end{equation}

Note that the eigenstates of the densitized triad operator are properly
normalized even though the eigen values span the real line. This is a
reflection of the fact that in loop quantization, only holonomies of
connections are well defined operators and {\em not} the connections
themselves.

As in LQG, the classical Hamiltonian constraint is promoted to a quantum
Hamiltonian operator using the volume operator whose action on the triad
basis states is given by
\begin{equation}
\hat{V} ~|\mu \rangle ~=~ {\left|\frac{1}{6}\gamma l_p^2 \mu \right|}^{\frac{3}{2}}
~|\mu \rangle   ~:=~ V_{\mu}~|\mu \rangle ~. 
\label{VolAction}
\end{equation}

In the isotropic context we have two classes to consider, namely
spatially flat and closed models. The quantization of the
corresponding Hamiltonian operators is given in \cite{IsoCosmo}.
By introducing a parameter $\eta$ we can deal both classes together. The
values $\eta = 0$ and $\eta = 1$ will give the flat and the closed
models respectively. 

In loop quantization of the classical gravitational Hamiltonian (in
connection formulation), one has to express the connections (and their
curvatures) in terms of appropriate holonomies of the connection. This
introduces a dimensionless, real parameter $\mu_0$ through a fiducial length
of the loops used in defining the holonomies. This parameter is to be
determined relating this symmetry reduced model to the full theory
\cite{Bohr}. For the purposes of this paper, it is simply treated as a
parameter of order 1.

The action of the gravitational Hamiltonian on the triad basis states is
then given by
\begin{eqnarray}
\hat{H}^{(\mu_0)}_{\text{grav}} |\mu \rangle & = & \left(\frac{3}{4
\kappa}\right)
{(\gamma^3 \mu_0^3 l_p^2)}^{-1} (V_{\mu+\mu_0} - V_{\mu-\mu_0})
\nonumber \\
& & \hspace{1.2cm} \left( ~e^{-i \mu_0\eta }|\mu + 4 \mu_0 \rangle ~-~ 
(2 + 4 \mu_0^2 \gamma^2\eta )|\mu \rangle ~+~ 
e^{i \mu_0\eta }|\mu - 4 \mu_0 \rangle ~\right)~ . 
\label{HGravAction}
\end{eqnarray}

Notice that the Hamiltonian connects states differing in their labels by
$\pm 4 \mu_0$. This is a direct consequence of the necessity of using
holonomy operators in the quantization of the Hamiltonian operator and
is responsible for leading to a difference equation below. 

General kinematical states $|s\rangle$, in the triad basis have the form 
\begin{equation}
|s \rangle ~=~ \sum_{\mu \in \mathbb{R}}~s_{\mu}~ |\mu \rangle ~,
\label{KinState}
\end{equation}
where $\mu$ run over some {\em countable} subset of $\mathbb{R}$. 

The Hamiltonian constraint of the classical theory is promoted as a
condition to define physical states, i.e.,
\begin{equation}
(\hat{H}^{(\mu_0)}_{\text{grav}} + \hat{H}^{(\mu_0)}_{\text{matter}}) |s \rangle ~=~ 0~.
\label{HamCons}
\end{equation}

In terms of $\tilde{s}_{\mu}:= e^{\frac{i \mu}{4}\eta} s_{\mu}$ and
The Hamiltonian constraint (\ref{HamCons}) translates into a difference
equation, 
\begin{eqnarray} \label{DifferenceEqn}
0 & = & A_{\mu + 4 \mu_0} \tilde{s}_{\mu + 4 \mu_0} ~-~ 
(2 + 4 \mu_0^2 \gamma^2 \eta) A_{\mu} \tilde{s}_{\mu} ~+~ 
A_{\mu - 4 \mu_0} \tilde{s}_{\mu - 4 \mu_0} \nonumber \\
& & \hspace{0.0cm} +~ 8 \kappa \gamma^2 {\mu_0}^3 
{\left(\frac{1}{6} \gamma l_p^2\right)}^{-\frac{1}{2}}
 H_{m}(\mu) \tilde{s}_{\mu} ~~~,~~~~~ \forall ~ \mu ~ \in \mathbb{R} \\
A_{\mu} & := & {|\mu + \mu_0|}^{\frac{3}{2}} -
{|\mu-\mu_0|}^{\frac{3}{2}}~~, \hspace{1.0cm} 
\hat{H}^{(\mu_0)}_{\text{matter}} |\mu\rangle ~ := ~ 
H_{m}(\mu) |\mu\rangle ~~ . \nonumber
\end{eqnarray}

$H_{m}(\mu)$ is a symbolic eigen value and we have assumed that
the matter couples to the gravity via the metric component and {\em not}
through the curvature component. In particular, $H_{m}(\mu = 0) =
0$. A couple of points are note worthy. 

Although $\mu$ takes all possible real values, the equation connects the
$\tilde{s}_{\mu}$ coefficients only in steps of $4 \mu_0$ making it a
difference equation for the coefficients. By putting $\mu := \nu +
(4\mu_0) n,~ n \in \mathbb{Z}, ~\nu \in (0, 4 \mu_0)$, one can see that
one has a continuous infinity of independent solutions of the difference
equation, labeled by $\nu, ~ S_n(\nu) := \tilde{s}_{\nu + 4 \mu_0 n}$.
For each $\nu$ an infinity of coefficients, $S_n(\nu), \forall n \in
\mathbb{Z}$, are determined by 2 `initial conditions' since the order of
the difference equation in terms of these coefficients is 2.
Coefficients belonging to different $\nu$ are mutually decoupled.
Denoting any two (fixed) independent solutions of the difference
equation by $\rho_n(\nu)$ and $\sigma_n(\nu)$, a general solution of the
second order difference equation, for {\em each $\nu \in (0, 4\mu_0)$}
can be expressed as $S_n(\nu) = S_0(\nu) \rho_n(\nu) + S_1(\nu)
\sigma_n(\nu)$ where $S_0(\nu), S_1(\nu)$ are arbitrary complex numbers.
Linearity of the equation implies that only their ratio is relevant and
for $\nu = 0$, this ratio is fixed \cite{EffHamiltonian}. A {\em general
solution} of the fundamental equation (\ref{DifferenceEqn}) is given by
a sum of the $S_n(\nu)$ solutions for a countable set of values of $\nu
\in [0, 4 \mu_0)$. Note that there are infinitely many ways of selecting
the countable subsets of values of $\nu$.  Since the coefficients
$A_{\mu}$ and the symbolic eigenvalues $H_{m}(\mu)$, both vanish for
$\mu = 0$, the coefficient $\tilde{s}_0$ {\em decouples} from all other
coefficients ($\tilde{s}_0$ of course occurs in the $\nu = 0$ sector
\cite{DynIn}). As the $\tilde{s}_{\mu}$ are defined for all $\mu \in
\mathbb{R}$, there is no break down of the dynamics at zero volume and
in this sense, the LQC dynamics is singularity free \cite{Sing}. 

For large values of $\mu \gg 4 \mu_0$ ($n \gg 1$), which correspond to
large volume, the coefficients $A_{\mu}$ become almost constant (up to a
common factor of $\sqrt{n}$) and the matter contribution is also
expected similarly to be almost constant. One then expects the
coefficients to vary slowly as $n$ is varied which in turn implies that
$S_n(\nu)$ also vary slowly with $n$. This suggests interpolating these
slowly varying {\em sequences} of coefficients by slowly varying,
sufficiently differentiable, {\em functions} of the continuous variable
$p(n) := \Case{1}{6} \gamma \lP^2 n$ \cite{SemiClass}. Using Taylor
expansion of the interpolating function, the difference equation for
$S_n(\nu)$ then implies a {\em differential} equation for the
interpolating function. This is referred to as a {\em continuum
approximation}. The terms up to (and including) second order
derivatives, turn out to be {\em independent} of $\gamma$ and the
differential equation, truncated to keep only these terms, matches with
the usual Wheeler--DeWitt equation of quantum cosmology.  This will be
referred to as a {\em pre-classical approximation}
\cite{FundamentalDisc}.

In the second step, one makes a WKB ansatz for solutions of the
differential equation. To the leading order in $\hbar$, one obtains a
Hamilton-Jacobi equation for the phase from which the effective
Hamiltonian is read-off.  In \cite{EffHamiltonian}, such an effective
classical Hamiltonian was obtained from the pre-classical approximation
and it was shown that the classical dynamics implied by the effective
Hamiltonian is also singularity free due to a generic occurrence of a
bounce \cite{GenBounce}. The derivation of the effective Hamiltonian was
based on a WKB approximation for the second order differential equation.
It is natural to seek to improve the pre-classical approximation by
including higher derivative terms in the differential equation. 

In this work, however, we will follow a different route. We will make
the WKB ansatz at the level of the difference equation itself. The
slowly varying property of the interpolating function will be applied to
the amplitude and the phase of the interpolating function. It is then
very easy to obtain an effective Hamiltonian with nontrivial dependence
on the Barbero-Immirzi parameter $\gamma$ and also the domain of
validity of WKB approximation. In the limit $\gamma \to 0$, one will
recover the earlier results of \cite{EffHamiltonian}.

\section{Derivation of Effective Hamiltonian}
\label{MainSectionI}
In anticipation of making contact with a classical description,
introduce the dimensionful variable $p(\mu) := \Case{1}{6} \gamma \lP^2
\mu $ as the continuous variable and an interpolating function $\psi(p)$
via $\psi(p(\mu)) := \tilde{s}_{\mu}$. Correspondingly, define $p_0 :=
\Case{1}{6} \gamma \lP^2 \mu_0$ which provides a convenient {\em scale}
to demarcate different regimes in $p$. 

Defining $A(p) := (\Case{1}{6} \gamma \lP^2)^{\Case{3}{2}} A_{\mu}, \ q
:= 4 p_0$ and replacing $\tilde{s}_{\mu}$ by the interpolating function 
$\psi(p)$, the fundamental equation (\ref{DifferenceEqn}) becomes, 
\begin{eqnarray}
0 & = & A(p + q)\psi(p + q) - \left(2 + 9
\frac{q^2}{\lP^4}\eta\right)A(p) \psi(p) + A(p - q)\psi(p - q)
\nonumber \\
& & ~ + \frac{9}{2} \frac{q^3}{\lP^4}\kappa H_m(p)\psi(p)
\label{FunctionalEqn}
\end{eqnarray}

At this stage one could use $\psi(p + q) = \sum_{k = 0}^{\infty}
\frac{q^k}{k!}\psi_k(p)$, with $\psi_k(p)$ denoting the $k^{th}$ derivative
of the function with respect to $p$, derive an infinite order
differential equation, consider WKB approximation to extract the
$o(\hbar)$ terms which gives a Hamilton-Jacobi equation and read-off the
effective Hamiltonian. Indeed, this is how we first arrived at the
effective Hamiltonian, but a much shorter route is available. 

The equation (\ref{FunctionalEqn}) can be thought of as a {\em
functional equation} determining $\psi(p)$. and of course it has
infinitely many solutions, corresponding to $\{S_n(\nu)\}, \nu \in [0,
4 \mu_0)$. So far there has been no approximation, we have only used
$p(\mu)$ instead of $\mu$ and $\psi(p)$ instead of $\tilde{s}_{\mu}$.

One can always write the complex function $\psi(p) := C(p)e^{i
\Case{\Phi(p)}{\hbar}}$. In the spirit of a continuum description, we
assume that the amplitude and the phase are Taylor expandable and write,
\begin{eqnarray}
C(p \pm q) & := & C(p)( \delta C_+ \pm \delta C_- )
~~~,~~~\Phi(p \pm q) ~ := ~ \Phi(p) + \delta \Phi_+ \pm \delta \Phi_-
\label{TaylorExpanssion} \\
\delta C_+ & := & \sum_{n = 0}^{\infty} \frac{q^{2n}}{(2 n) !}
\frac{C_{2n}(p)}{C(p)} ~~~,~~~
\delta C_- ~ := ~ \sum_{n = 0}^{\infty} \frac{q^{2n + 1}}{(2 n + 1)!}
\frac{C_{2n + 1}(p)}{C(p)} \label{DeltaAmplitude} \\
\delta \Phi_+ & := & \sum_{n = 1}^{\infty} \frac{q^{2n}}{(2 n) !}
\Phi_{2n}(p) ~~~,~~~
\delta \Phi_- ~ := ~ \sum_{n = 0}^{\infty} \frac{q^{2n + 1}}{(2 n + 1)!}
\Phi_{2n + 1}(p) \label{DeltaPhase} %\\
\end{eqnarray}

We have just separated the even and odd number of derivative terms for
later convenience. Substitution of $\psi(p \pm q) ~=~ C(p \pm
q)e^{\frac{i}{\hbar}\Phi(p \pm q)}$ in (\ref{FunctionalEqn}) leads to,
\begin{eqnarray}\label{WKBForm}
0 & = & H_m - \frac{1}{\kappa} \frac{6}{\epsilon^3 \lP^2} ( 2 + \epsilon^2
\eta ) A(p) +  \frac{1}{\kappa} \frac{6}{\epsilon^3 \lP^2} \times \nonumber \\
& & 
\left[ (B_+(p, q) \delta C_+(p) + B_-(p, q) \delta C_-(p))\left(
\mathrm{cos}\frac{\delta \Phi_+}{\hbar} \ 
\mathrm{cos}\frac{\delta \Phi_-}{\hbar} \right) \right. \nonumber \\
& & 
- (B_-(p, q) \delta C_+(p) + B_+(p, q) \delta C_-(p))\left(
\mathrm{sin}\frac{\delta \Phi_+}{\hbar} \ 
\mathrm{sin}\frac{\delta \Phi_-}{\hbar} \right) \nonumber \\
& & 
+ i (B_-(p, q) \delta C_+(p) + B_+(p, q) \delta C_-(p))\left(
\mathrm{cos}\frac{\delta \Phi_+}{\hbar} \ 
\mathrm{sin}\frac{\delta \Phi_-}{\hbar} \right) \nonumber \\
& & \left.
+ i (B_+(p, q) \delta C_+(p) + B_-(p, q) \delta C_-(p))\left(
\mathrm{sin}\frac{\delta \Phi_+}{\hbar} \ 
\mathrm{cos}\frac{\delta \Phi_-}{\hbar} \right) \right] \nonumber \\
B_{\pm}(p, q) & := & A(p + q) \pm A(p - q)
\end{eqnarray}

Thus, one has {\em two equations} for the {\em four combinations},
$\delta C_{\pm}, \delta \Phi_{\pm}$. 

Now we assume that the amplitude and the phase are {\em slowly varying}
functions of $p$ over a range $q$ i.e.  when compared over a range $\pm
q$, successive terms of a Taylor expansion about $p$ are smaller than
the preceeding terms.  For example, $|\Case{q C'}{C}| \ll 1, |\Case{q^2
C''}{2 q C'}| \ll 1$ etc. and similarly for the phase.

For such slowly varying solutions of (\ref{FunctionalEqn}), we can
approximate the equation by keeping only the first non-trivial
derivatives i.e. $\delta C_+ = 1,\  \delta C_- = \Case{q C'}{C}$ and
$\delta \Phi_- = q \Phi',\  \delta \Phi_+ = \Case{q^2 \Phi''}{2}$.  To
arrive at a Hamilton-Jacobi equation, we identify $\Phi'(p) :=
\Case{3}{\kappa} K$ which implies that $\Case{q \Phi'}{\hbar} = \Case{2
\mu_0 \gamma \lP^2 \Phi'}{3 \hbar} = \epsilon K$, where we have used the
definitions $\epsilon = 2 \mu_0 \gamma$ and $\lP^2 = \kappa \hbar$.
This combination of the first derivative of the phase thus has no
explicit dependence on $\hbar$. 

The terms in the real and imaginary equations can be organized according
to powers of $\hbar$. For sine and cosine of $\delta \Phi_+$, we have to
use the power series expansions and keep only the leading powers of
$\hbar$. For sine and cosine of $\delta \Phi_-$, no expansion is needed
since there is no $\hbar$ dependence. The real and imaginary equations
can then be written as, 
\begin{eqnarray} 
0 & = & H_m - \frac{1}{\kappa} \frac{6}{\epsilon^3
\lP^2} (2 + \epsilon^2 \eta) A(p) + %\nonumber \\
%& &
\frac{1}{\kappa} \frac{6}{\epsilon^3 \lP^2} \left[ \left\{ 1 +
\frac{B_-(p,q)}{B_+(p,q)} \frac{q C'}{C} \right\} B_+(p, q)\
\mathrm{cos}(\epsilon K) \right. \nonumber \\ & & \left. \hspace{1.0cm}
- B_+(p,q)\left\{\frac{B_-(p,q)}{B_+(p,q)} + \frac{q
C'}{C}\right\}\left(\frac{q^2 \Phi''}{2 \hbar}\right)
\mathrm{sin}\epsilon K \right] \label{RealEqn}\\
0 & = & \left(\frac{q^2 \Phi''(p)}{2 \hbar \epsilon K} \right)\left\{1 +
\frac{B_-(p, q)}{B_+(p, q)}\frac{q C'}{C}\right\} + \left(\frac{B_-(p,
q)}{B_+(p, q)} + \frac{q C'}{C} \right) \
\frac{\mathrm{tan}\left(\epsilon K \right)}{\epsilon K} \label{ImEqn}
\end{eqnarray}

On physical grounds, one expects a classical approximation ($\hbar^0$
terms), to be valid only for scales larger than the quantum geometry
scale set by $q$ and therefore we limit to the regime $p \ge q$. In this
regime, the coefficients $A(p), B_{\pm}(p,q)$ behave as
\cite{EffHamiltonian},
\begin{eqnarray}
A(p) ~ & \approx & ~ \frac{3}{4} q \sqrt{p} - o(p^{-\Case{3}{2}}) \nonumber \\
B_+(p, q) ~ & \approx & ~ \frac{3}{2} q \sqrt{p} - o(p^{-\Case{3}{2}})
\nonumber \\
B_-(p, q) ~ & \approx & ~ \frac{3}{4} q^2 \ p^{- \frac{1}{2}} 
+ o(p^{-\frac{5}{2}}) \label{LargeVolLim}
\end{eqnarray}

Noting that $q = \Case{1}{3} \epsilon \lP^2$ and keeping only the leading 
powers of $\lP^2$ (or $\hbar$), the real and imaginary equations become
(WKB approximation),
\begin{eqnarray} 
0 & = & H_m - \frac{1}{\kappa}
\frac{6}{\epsilon^3 \lP^2} (2 + \epsilon^2 \eta) A(p) + %\nonumber \\
%& &
\frac{1}{\kappa} \frac{6}{\epsilon^3 \lP^2} \left[ B_+(p, q)
\mathrm{cos}(\epsilon K) \right] \label{CorrectHamJac}\\
0 & = & \left(\frac{q^2 \Phi''(p)}{2 \hbar \epsilon K} \right) +
\left(\frac{B_-(p, q)}{B_+(p, q)} + \frac{q C'}{C} \right) \
\frac{\mathrm{tan}\left(\epsilon K \right)}{\epsilon K}
\label{WKBCondition} 
\end{eqnarray}

The real equation, (\ref{CorrectHamJac}) is of $o(\hbar^0)$ and is a
Hamilton-Jacobi equation for the phase. The right hand side, viewed as a
function of $p, K$, it is the {\em effective Hamiltonian constraint}.
The imaginary equation, (\ref{WKBCondition}) is of $o(\hbar)$ and is a
differential equation for the amplitude, given the phase determined by
the real equation. The equations for the phase and the amplitude are
decoupled. For self consistency of the WKB approximation, the solutions
have to be slowly varying. In general, the solutions will be slowly
varying only over some intervals along the $p$-axis and such interval(s)
will be the domain of validity of the WKB approximation. One can infer
the domain of validity as follows.

Consider the eqn. (\ref{WKBCondition}). By definition of slowly varying
phase, the absolute value of the first term must be much smaller than 1.
Since smallest value of $\Case{\text{tan}(\epsilon K)}{\epsilon K}$ is
1, (the absolute value of) the bracket in the second term must be
smaller than 1. Since $\Case{q C'}{C}$ is also small by definition of
slowly varying amplitude, we must have $|\Case{B_-}{B_+}| \le 1$. This
immediately requires $p \ge q = \Case{\epsilon\lP^2}{3}$ and is
obviously true in the regime under consideration. Evidently, for {\em
smaller volumes}, $p \gtrsim q$, where $\Case{B_-}{B_+}$ is larger, we
must have $\Case{\text{tan}\epsilon K}{\epsilon K} \gtrsim 1$ i.e.
$\epsilon K \approx 0$. By contrast, for {\em larger volumes}, $p \gg
q$, larger values of $\epsilon K ( < \Case{\pi}{2} )$ are permitted.
Thus, the equation (\ref{WKBCondition}) serves to identify a region of
the classical {\em phase space} where the effective classical
description is valid. The domain of validity of effective Hamiltonian
and its relation to the usual general relativity (GR) Hamiltonian is
discussed further in the remarks below.

It is convenient to write the real equation in the form,
\begin{eqnarray} 
0 & = & - \frac{1}{\kappa}\left[
\left(\frac{3}{\epsilon\lP^2}\right) \left\{ B_+(p, q)
\left(\frac{4}{\epsilon^2}\mathrm{sin}^2\left(\frac{\epsilon}{2}K
\right)\right) + 2 A(p) \eta \right\}\right] \nonumber \\
& & + \frac{1}{\kappa} \left[\left(\frac{6}{\epsilon^3\lP^2}\right)
\left\{B_+(p, q) - 2 A(p)\right\} \right] ~ + ~ H_m \label{HamJacEqn}% \\ 
\end{eqnarray}

The second square bracket in eq.(\ref{HamJacEqn}) is precisely the {\em
quantum geometry potential}, $W_{\text{qg}}/\kappa$,
\cite{EffHamiltonian}. It is unaffected by the implicit inclusion of all
terms of the Taylor expansion of $\psi(p)$.  Since the leading terms in
$B_+ - 2 A$ vanish, the quantum geometry potential term is independent
of $\epsilon$. But it is also higher order in $\hbar$.  From now on,
this term is suppressed.  It is obvious that in the limit $\epsilon \to
0$, one gets back the expression for the effective Hamiltonian
constraint obtained from the pre-classical approximation in
\cite{EffHamiltonian}. 

Let us focus on the effective Hamiltonian constraint in the large volume
regime. As mentioned earlier, the matter is assumed to couple only
through metric components and thus the matter Hamiltonian has dependence
on $p$ but not on $K$. For the moment, we will not be explicit about the
matter Hamiltonian and simply view it to be a function $p$ and matter
degrees of freedom symbolically denoted by $\phi, p_{\phi}$. In the next
section we will be more explicit about it.  The relevant Hamiltonian
constraint then becomes (suppressing the quantum geometry potential):
\begin{eqnarray} \label{HamConstraint}
0 & = & - \frac{3}{2 \kappa}\sqrt{p} \left[ 
\frac{4}{\epsilon^2}\mathrm{sin}^2\left(\frac{\epsilon}{2}K \right) + 
\eta \right] 
+ ~ H_m(p, \phi, p_{\phi})
%+ ~ \left[ \frac{1}{2} p^{- 3/2} p^2_{\phi} + p^{3/2} V(\phi) \right]
\end{eqnarray}

To see what the modified Hamiltonian constraint implies for cosmological
space-times, we have to obtain and solve the Hamilton's equations. To
this end, we identify $|p| = \Case{1}{4} a^2$ and choose the synchronous
time as the evolution parameter (Lapse = 1). It is straight forward to
obtain the Hamilton's equations of motion as:
\begin{eqnarray} \frac{d p}{d t} & = & - \frac{\sqrt{p}}{\epsilon}
\mathrm{sin}(\epsilon K) ~~~ = ~~\frac{a \dot{a}}{2} \label{PDotEqn}\\
\frac{d K}{d t} & = & - \frac{\kappa}{3} \frac{\partial H_m(\phi,
p_{\phi}, p)}{\partial p} + \frac{1}{4 \sqrt{p}} \left[
\frac{4}{\epsilon^2}\mathrm{sin}^2\left(\frac{\epsilon}{2}K \right) +
\eta \right] \label{KDotEqn}\\
%
%%\frac{d \phi}{d t} & = & p^{- 3/2} p_{\phi} \label{PhiDotEqn} \\
\frac{d \phi}{d t} & = & \frac{\partial H_m(p, \phi, p_{\phi})}
{\partial p_{\phi}} \label{PhiDotEqn} \\
%%\frac{d p_{\phi}}{d t} & = &  - p^{ 3/2}\frac{d V(\phi)}{d \phi}
\frac{d p_{\phi}}{d t} & = &  - \frac{\partial H_m(p, \phi, p_{\phi})}
{\partial \phi} \label{PPhiDotEqn} \end{eqnarray}

The equation (\ref{HamConstraint}) is the modification of the Friedmann
equation while (\ref{PDotEqn}, \ref{KDotEqn}) lead to modified
Raychoudhuri equation {\em after eliminating $K, p$ in favour of the
scale factor and its time derivatives.}  In particular, (\ref{PDotEqn})
gives $\dot{a} = - \Case{\mathrm{sin}(\epsilon K)}{\epsilon}$ and leads
to $\mathrm{cos}(\epsilon K) = \sqrt{1 - \epsilon^2 \dot{a}^2}$
($\epsilon \to 0$ limit fixes the sign of the square root).  Thus the
Hamiltonian constraint can be expressed in terms of $\dot{a}$.
Furthermore, since the constraint is obviously preserved along the
solutions of the Hamilton's equations, we can obtain $\dot{K}$ in terms
of $H_m$.  Now is it is straightforward to construct left hand sides of
the Friedmann and the Raychoudhuri equation and comparing with usual
Einstein equations, read-off the effective density and pressure. One
gets,
\begin{eqnarray} 
\frac{\kappa}{2}\rho_{\text{eff}} ~:= ~ 3 \left(
\frac{\dot{a}^2 + \eta}{a^2} \right) & = & \frac{\kappa}{2}\left[ 8
a^{-3} H_m - \frac{3 \epsilon^2}{2 \kappa a^2}\left( \frac{4 \kappa}{3
a} H_m - \eta\right)^2 \right] \label{FriedmannEqn} \\
- \frac{\kappa}{4}\left( \rho_{\text{eff}} + 3 P_{\text{eff}}\right)
~:=~ 3 \frac{\ddot{a}}{a} & = & - \frac{\kappa}{4} \left[ 8 a^{-3}( H_m
- a \frac{\partial H_m}{\partial a} ) \left\{ 1 -
\frac{\epsilon^2}{2}\left( \frac{4 \kappa}{3 a} H_m  - \eta
\right)\right\} \right] \label{RaychoudhuriEqn} 
\end{eqnarray}
For future reference, we note that the effective density and pressure
can be expressed as \footnote{Our normalizations are inherited from the
normalizations used in Bianchi models, by equating the three triad
variables $|p_I|$ to the triad variable for the isotropic specialization
$|p|$ which equals $\Case{a^2}{4}$. The Bianchi normalized volume thus
equals $\Case{a^3}{8}$. A Hamiltonian is thus $\Case{a^3}{8}$ times the
Hamiltonian density \cite{EffHamiltonian}.}
\begin{eqnarray} 
\rho_{\text{eff}} & = & 8 a^{-3} \bar{H}_m \nonumber \\
\left( \rho_{\text{eff}} + 3 P_{\text{eff}}\right)
& = & 8 a^{-3}( \bar{H}_m - a \frac{\partial \bar{H}_m}{\partial a} )
\label{Effpressure} \hspace{3.0cm}\text{where,}\nonumber \\ 
\bar{H}_m & := & H_m - \frac{3 \epsilon^2 a}{16 \kappa}\left(\frac{4
\kappa}{3 a} H_m - \eta\right)^2 \label{HBarDef}
\end{eqnarray}

As $\epsilon \to 0\ (\gamma \to 0)$, the right hand sides of the above
equations go over to the equations of \cite{EffHamiltonian} and so do
the effective density and effective pressure, with the quantum geometry
potential terms suppressed. 

The advantage now is that the Raychoudhuri equation is automatically
satisfied once the Hamilton's equations for the matter hold and the
Friedmann equation holds. Thus we have the $\epsilon$-corrected, three
coupled (for a single matter degree of freedom), first order ordinary
differential equations which go over to the classical equations when
$\epsilon \to 0$. It is now a simple task to compare the solutions
with/without $\epsilon$ corrections, for the same initial conditions
consistent with small $\dot{a}$ and large $a$.  Some solutions are
discussed in the next section. 

Several remarks are in order:

(1) It is somewhat surprising that the parameters related to quantum
geometry, $\mu_0$ and $\gamma$ (which have to be non-zero), appear in
the effective Hamiltonian which is {\em independent} of $\hbar$ i.e., in
a {\em classical} description.

To trace how this happens, recall that the basic phase space variables
in isotropic LQC (in the connection formulation) are $c, p$ and the
conjugate momentum variable $K$ is related to the connection variable
$c$ by $c = \gamma K$ (for spatially flat model for definiteness). In
loop quantizing the Hamiltonian constraint, the classical constraint is
to be expressed in terms of the holonomies of the connection (equation
(34) of \cite{Bohr}). In this process, the parameter $\mu_0$ enters.
The classical constraint so obtained, precisely contains the
$\text{sin}^2 \mu_0 c$ which is the same as $\text{sin}^2 \Case{\epsilon
K}{2}$. The parameter $\epsilon$ is thus present already at the
classical level ($\hbar^0$). Within a strictly classical context, one
can view it as a ``regulator" and remove it at will, by taking $\epsilon
\to 0$ to recover the classical Einstein Hamiltonian.  Within a strictly
WKB context also one can remove it since the corrections to WKB are
positive powers of $\Case{q}{p} \sim o(\epsilon)$. However, from the
perspective of the exact loop quantization, we {\em cannot} take
$\epsilon \to 0$ \cite{Bohr}. While viewing the WKB solutions as an
approximation to the exact LQC solutions, we retain the link to the
exact solutions by keeping $\epsilon \ne 0$.

(2) The classical {\em GR Hamiltonian} is the expression obtained from the
effective Hamiltonian in the limit $\epsilon K \to 0$ (and large
volume). This could be achieved by {\em either} taking $\epsilon \to 0,
K$ fixed, which is a natural classical GR perspective since the
classical GR has no $\epsilon$ {\em or} by taking $K \to 0, \epsilon$
fixed which is the appropriate quantum perspective in LQC. This opens up
the possibility that there could be modifications of {\em classical} GR
especially when the conjugate momentum ($K$) gets somewhat larger. As
noted already, for smaller volumes, the domain of WKB approximation
restricted by equation (\ref{WKBCondition}), requires $K$ to be small
and the effective Hamiltonian reduces to the GR Hamiltonian. For small
$K$, $K \approx - \dot{a}$ and corresponds to the extrinsic curvature (a
geometrical quantity). For larger values of $K$, the relation between
$K$ and the extrinsic curvature is given by equation (\ref{PDotEqn}). 

(3) Since for smaller volumes, the effective Hamiltonian goes over to
the GR Hamiltonian, the effective density and pressure defined in
(\ref{FriedmannEqn}, \ref{RaychoudhuriEqn}) go over to those
defined in \cite{EffHamiltonian}. Consequently, the genericness of
inflation \cite{GenInflation} is insensitive to the $\epsilon$
parameter. Furthermore, since the quantum geometry potential is 
unaffected, the genericness of bounce also continues to hold (in this
regime, the quantum geometry potential must be retained in equation
(\ref{HamJacEqn})) \cite{GenBounce}.  

(4) The occurrence of the trigonometric function of $K$ immediately
implies two bounds. The effective Hamiltonian constraint,
(\ref{HamConstraint}), implies that is that the matter Hamiltonian is
necessarily bounded: $\eta \le \Case{4 \kappa}{3} a^{-1} H_m \le \eta +
\Case{4}{\epsilon^2}$ or equivalently, $0 \leq \Case{4 \kappa}{3} a^{-1} H_m -
\eta \le \Case{4}{\epsilon^2}$. Equation (\ref{PDotEqn}) on the other
hand implies a bound on the expansion rate: $|\dot{a}| \le \epsilon^{-1}$.

Consider the bound on the matter Hamiltonian. Momentarily, let us write
$H_m = D \Case{a^3}{8}$ where the `density' $D$ may include
modifications from inverse triad operator \cite{InvScale}. We have the
two regimes of large and small volumes, two possibilities for spatial
curvature, $\eta$ and two bounds on the matter Hamiltonian to consider.
For {\em large volumes}, we may identify $D$ with the usual energy
densities for dust and radiation (late time dominant matter) or with a
positive cosmological constant. For dust and radiation, the upper bound
is satisfied for both flat and closed models but the lower bound will be
violated for closed models ($\eta = 1$) and hence a re-collapse must
commence.  If $D$ is identified with a positive cosmological constant,
then the lower bound will be satisfied but the upper bound {\em will} be
violated for {\em large} enough volumes irrespective of $\eta$, resulting
in re-collapse. For {\em smaller volumes}, cosmological constant will
satisfy the upper bound but will violate the lower bound for a close
model implying a bounce. In this regime, the dust/radiation will
potentially violate the upper bound. However, for smaller volumes, $D$
has modifications from the inverse volume operator which imply that $D$
goes to a constant ensuring that the upper bound is indeed satisfied. In
the small volume regime, one also has a contribution from the quantum
geometry potential which implies that irrespective of $\eta$, the lower
bound will be violated implying a bounce \cite{GenBounce}.  Since the
effective constraint excludes evolutions which violate either of the
bounds, a violation in the large volume regime must encounter a
re-collapse while a violation in the small volume regime indicates a
bounce. The solutions discussed in the next section illustrate these
implications.

Consider now the bound on the expansion rate, $0 \leq \dot{a} \le
\epsilon^{-1}$. Let us trace a possible scenario beginning with a bounce
for a spatially flat model. Let us set $t = 0$ at the bounce time and
$a(0) := a_{\text{bounce}} \gtrsim \lP$. Near this time, we must include
the quantum geometry potential i.e. $H_m \to \tilde{H}_m := H_m +
\Case{W_{\text{qg}}}{\kappa}$ in the equation (\ref{FriedmannEqn}) and
$H_m \sim h a^3$, where $h$ is a constant \cite{GenInflation}. The
bounce is determined by $\tilde{H}_m = 0$ and gives $a_0 \sim h^{-1/6}$
\cite{GenBounce}. Close to $t \approx 0$, the scale factor will behave
as, $a(t) \approx a_{\text{bounce}} + \alpha t^{\beta}, \beta \ge 2$ and
$\alpha$ is a positive constant.  Since $\dot{a} = \alpha \beta t^{\beta
-1}$, the bound {\em will} be violated in a finite time. The bounce
phase will go over to an exponentially expanding phase when the quantum
geometry potential becomes negligible. This will happen well before
saturating the bound on $\dot{a}$, even if $a_{\text{bounce}}$ is just a
few times $\lP$.  Suppose the bounce phase ends at time $t_1$ and the
universe enters a phase of exponential expansion, $a(t) \sim e^{h t}$.
The exponent is determined by the parameters $\alpha, \beta$ and the
duration $t_1$ of the preceeding bounce phase.  Now $\dot{a}(t) = a(t)
h$ which increases monotonically.  This phase must again end in a finite
time (at $t_2$ say) to satisfy the bound.  Since
$\Case{\dot{a}(t_2)}{\dot{a}(t_1)} = \Case{a(t_2)}{a(t_1)}$ and
$\dot{a}(t_2) \le \epsilon^{-1}$, the maximum expansion is effectively
controlled by $\dot{a}(t_1)$.  Once again, the exponential inflation can
end before saturating the bound and these details will depend on the
precise matter dynamics (eg the form of scalar field potential). One can
repeat the scenario for power law accelerated expansion and have a
similar finite duration phase. If the bound gets saturated at any stage,
the acceleration must go over to deceleration possibly reaching a
re-collapse, contraction and another bounce and so on. If however the
inflationary phase ends prior to saturating the bound and the universe
enters a decelerated phase such as radiation domination followed by
matter domination, the bound will continue to hold. Note that such a
scenario is also conceivable without a bound on $\dot{a}$, however with
a bounded $\dot{a}$ the scenario becomes more plausible.  The bound is
also specified in terms of the discreteness parameter, $\epsilon$, which
is expected to be constrained by the full (inhomogeneous) loop quantum
gravity.

(5) Finally, let us note that $\dot{a}$ is the extrinsic curvature of
the symmetry adapted hypersurface (and is also the rate of change of the
physical volume of the universe) and is gauge invariant. The
modifications to GR due to the non-zero parameter $\epsilon$ are
manifested in the modified coupling (quadratic in matter Hamiltonian)
between matter and geometry. The GR domain, contained within the domain
of validity of WKB, can also be given as: $|p| = \Case{a^2}{4} \gg q =
\Case{\epsilon \lP^2}{3}$ translates into $a \gg
\sqrt{\Case{4\epsilon}{3}} \lP $ while $|\epsilon K| \ll \Case{\pi}{2}$
translates into $0 \leq \ \Case{4 \kappa}{3 a} H_m - \eta\ \ll \Case{\pi^2}{4
\epsilon^2} $. The effective Hamiltonian and subsequent analysis being
$o(\hbar^0)$, is insensitive to factor ordering in the Hamiltonian
constraint.

\section{Modified solutions}
\label{SolnSection}
Let us consider a few solutions of the modified equations. These are not
meant to be phenomenologically realistic solutions, but are to be viewed as
indicating modification to GR solutions. In particular
we consider the cases of a minimally coupled homogeneous scalar field,
phenomenological matter with a constant equation of state and positive
cosmological constant viewed as a special case of phenomenological
matter.

\subsection{Minimally coupled massive scalar field}
\label{MinScalar}

For simplicity, let us take the matter sector consisting of a scalar
field, $\phi$, minimally coupled to gravity. Then its usual classical
Hamiltonian is given by $H_m(\phi, p_{\phi}, p) = \Case{1}{2}p^{-3/2}
p_{\phi}^2 + p^{3/2} V(\phi)$. Its quantization involves two parts --
(a) quantization of $p^{- \Case{3}{2}}$ operator and (b) Bohr
quantization of the scalar field itself. For large volume, the inverse
volume operator in the triad representation just goes over to the
classical expression. Its correction terms involve higher powers of
$\hbar$ (the terms suppressed by inverse powers of $p$ in the
coefficients $B_{+}(p, q), A(p, q)$, likewise involve higher powers of
$\hbar$). Thus, to $o(\hbar^0)$, these modifications are irrelevant in
the large volume regime. The Bohr quantization appears not to introduce
$\gamma-$corrections to the usual classical matter Hamiltonian. In this
paper we will pretend that the matter is quantized in the usual
Schrodinger quantization. Loop quantization of matter with or without
non-minimal coupling (to mimic a dilaton) will be presented elsewhere.

The solutions are obtained numerically with $V(\phi) := \Case{1}{2}m^2
\phi^2$ and with initial conditions chosen to indicate different types of
behaviours. We use geometrized units ($\kappa = 1$ and speed of light
equal to 1). Introducing an arbitrary length scale $\bar{a}$, various
quantities has following dimensions.

\begin{equation}\label{Dimensions}
H_m \sim \bar{a}~,~ p_{\phi} \sim \bar{a}^2~,~\phi \sim \bar{a}^0~,~m
\sim \bar{a}^{-1}~,~t \sim \bar{a} \ .
\end{equation}

Scaling the quantities by the appropriate power of $\bar{a}$ all
equations are rendered dimensionless and integrated numerically. Since
we use the usual classical form of the matter Hamiltonian without the
corrections from the inverse volume, the length scale $\bar{a}$ is
suitably large (eg $\bar{a} \gtrsim 100 \lP$). 

\begin{figure}[htb]
\begin{center}
\includegraphics[width=15.0cm]{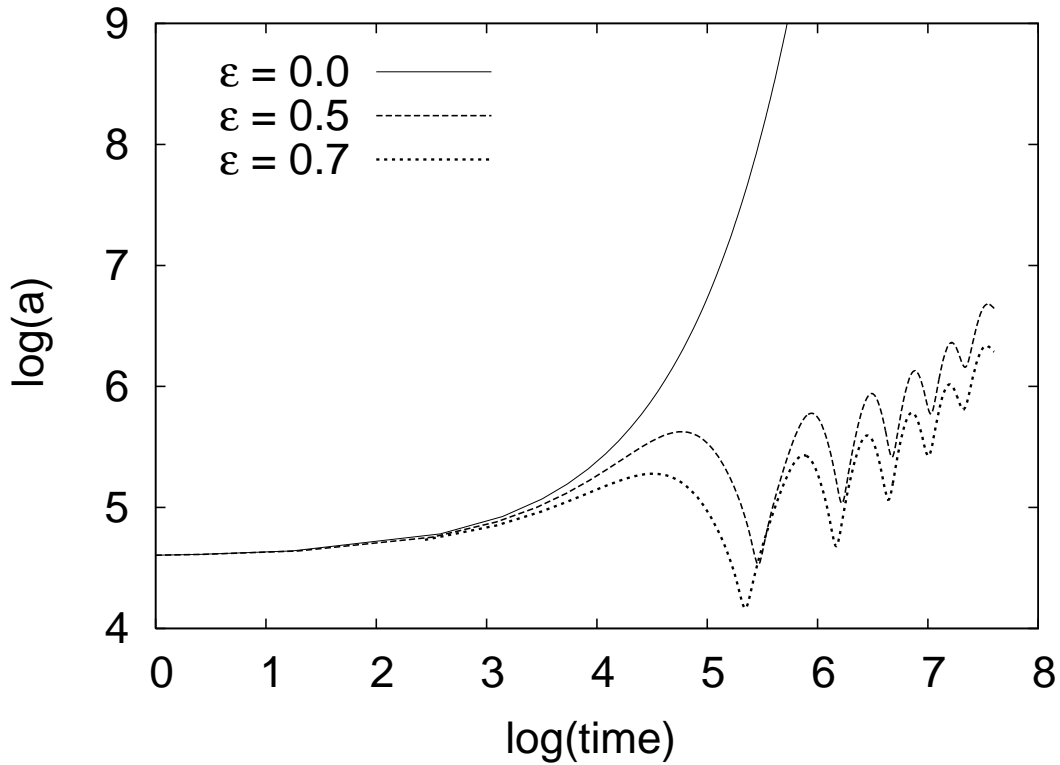}
\end{center}
\caption{The initial values here are: $\phi = 50, p_{\phi} = 50$. For late
times the evolutions show multiple bounces and re-collapses for non-zero
$\epsilon$.} \label{FiftyPhi}
\end{figure}

For the plots shown below in figure(\ref{OnePhi}), figure(\ref{TenPhi})
and figure(\ref{FiftyPhi}), the dimensionless scale factor assumed to be
100 and the dimensionless mass is taken to be 0.001. Three different
values of the discreteness parameter are chosen: $\epsilon = 0.0,
0.5, 0.7$.  Three different initial conditions for the scalar field and
its momentum are taken namely $(\phi, p_{\phi}) = (1, 10000), (10,
10000), (50, 50)$. As can be expected, the initial value of the scalar
field has stronger effect on the evolution. For smaller value of the
scalar field, the effect of non-zero $\epsilon$ is virtually absent
while for larger values one sees the multiple re-collapse/bounce
possibilities. For the plots, spatially flat model is considered ($\eta
= 0$). Only the evolution of the scale factor is shown in the figures.
The scalar field typically increases first and then decreases to smaller
values. Both the increase and the decrease of the scalar field are
steeper for larger values of $\epsilon$. The scalar field momentum also
shows similar behaviour. The decrease however is much sharper and occurs
at {\em later} times for larger values of $\epsilon$.

\begin{figure}[htb]
\begin{center}
\includegraphics[width=15.0cm]{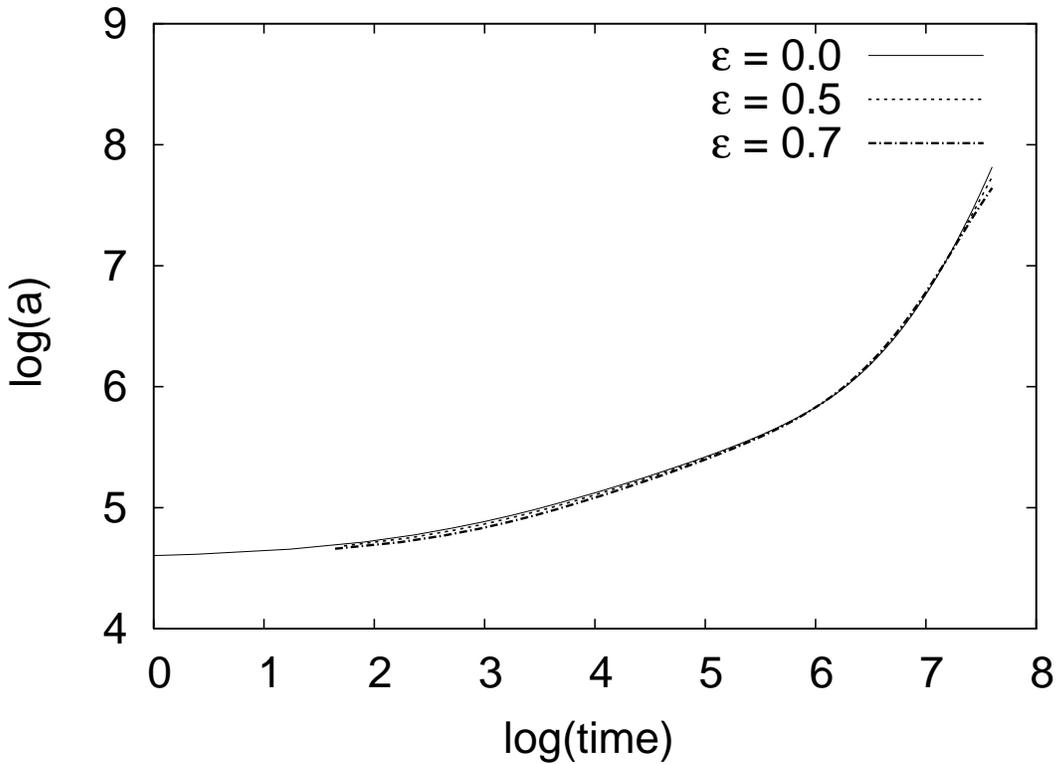}
\end{center}
\caption{The plot is for the initial values $\phi = 1, p_{\phi} = 10000$.
The discreteness corrections are very small.} \label{OnePhi}
\end{figure}

\begin{figure}[htb]
\begin{center}
\includegraphics[width=15.0cm]{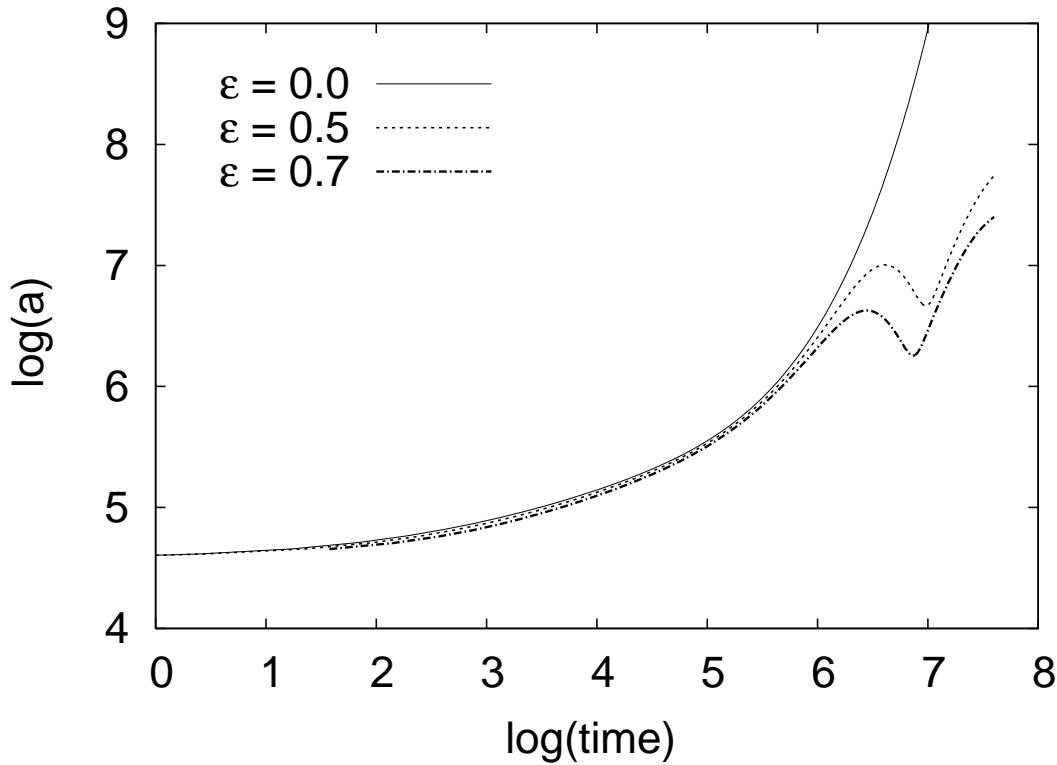}
\end{center}
\caption{This plot is for the initial values $\phi = 10, p_{\phi} = 10000$.
For late times, deviations due to non-zero $\epsilon$ are clearly visible.}
\label{TenPhi}
\end{figure}

\subsection{Phenomenological Matter}
In fact, it is rather easy to consider more general case of
phenomenological matter if one proceeds somewhat naively. By this we
mean the following. 

In the current treatment, the matter is also supposed to be described by
a Hamiltonian. However, for the usual `dust' and `radiation' which are
supposed to represent non-relativistic and relativistic matter
respectively, one does not write a Hamiltonian. The underlying matter
dynamics (with time scale shorter than that of expansion of the
universe) is supposed to ensure a thermal equilibrium. Such matter is
then thermodynamically described in terms of an energy density, $\rho$
and pressure, $P$, with the equilibrium property implying an equation of
state, $P = \omega \rho$. The first law of thermodynamics in conjunction
with adiabatic expansion implies the `continuity equation' $a \Case{d
\rho}{d a} = -3 \rho (1 + \omega)$ which can be integrated to give,
\begin{equation} 
\rho(a) ~=~ \bar{\rho}e^{-3 \int_{\bar{a}}^a \left\{ 1
+ \omega(a') \right\} \frac{da'}{a'} } ~~~=~~ \bar{\rho}
\left(\frac{a}{\bar{a}}\right)^{-3(1 + \omega)} ~,~~~{\mbox{when
$\omega$ is constant.}} 
\end{equation}

The equilibrium property also implies that the density and pressure are
homogeneous. Such matter in (approximate) equilibrium is coupled to
gravity via a stress tensor of the form of the stress tensor of a
prefect fluid (and the matter is correspondingly referred to as a
perfect fluid) which is of course the most general form of stress tensor
in the context of homogeneity and isotropy. This stress tensor is
conserved by virtue of the continuity equation implied by adiabatic
expansion and the first law of thermodynamics. This way of coupling matter
to gravity is thus consistent with the Einstein equation.

It is conceivable that some of the mechanisms (eg microscopic matter
dynamics) involved in establishing thermal equilibrium may be modified,
especially if the expansion time scale becomes comparable to that of the
matter processes responsible for establishing thermal equilibrium.
However, for the large volume regime we are considering, one may
justifiably assume that the microscopic matter dynamics are unchanged
and lead to thermal equilibrium with the usual equations of states. If
this assumption is granted then such sources can be incorporated by
taking $H_m(a) := \rho(a) \frac{a^3}{8}$, with the scale factor
dependence of matter Hamiltonian explicitly specified via that of the
thermodynamical energy density, $\rho(a)$. In view of the expressions in
(\ref{HBarDef}), this is equivalent to putting $\bar{H}_m =
\rho_{\text{eff}} \frac{a^3}{8}$. The assumption made above amounts to
viewing the effective equation of state, $\omega_{\text{eff}} :=
P_{\text{eff}}/\rho_{\text{eff}}$ being determined from the usual
equation of state, $\omega$. The effective density and pressure are thus
not ascribed any thermodynamic origin, but are viewed as prescribing a
modified coupling of the matter to gravity.

The evolution of the scale factor is now obtained by solving just the
(modified) Friedmann equation (\ref{FriedmannEqn}) for various choices
of $\omega(a)$. For example, for constant equation of state, $\omega$,
and for expanding universe, 
\begin{equation}
\dot{a} ~=~ + \left[ \left\{ {\frac{\kappa \bar{\rho}\bar{a}^2}{6}} 
\left(\frac{a}{\bar{a}}\right)^{-1 - 3 \omega} - \eta\right\} - 
\frac{\epsilon^2}{4}\left\{ {\frac{\kappa \bar{\rho}\bar{a}^2}{6}}
\left(\frac{a}{\bar{a}}\right)^{-1 - 3 \omega} - \eta\right\}^2 \right] 
^{\frac{1}{2}}
\end{equation}

Introducing dimensionless quantities,
\begin{equation}\label{DimlessDefns}
\sigma^2 ~ := ~ \frac{\kappa \bar{\rho}\bar{a}^2}{6} ~~,~~ \xi ~ := ~
\frac{a}{\bar{a}} ~~,~~ \tau ~ := ~ \frac{t}{\bar{a}}~~,~~ ' ~ :=~
\frac{d}{d \tau} \ ,
\end{equation}
the Friedmann equation can be written as,
\begin{eqnarray}
\xi' & = & \sigma \xi^{-(1 + 3 \omega)} \sqrt{ \xi^{(1 + 3 \omega)} -
\left(\frac{\sigma \epsilon}{2}\right)^2 } {\mbox{\hspace{3.0cm}, ~~ 
(flat model)}} \label{FlatModel} \\
\xi' & = & \sqrt{ \sigma^2 \xi^{-(1 + 3 \omega)} - 1} \sqrt{ 1 -
\frac{\epsilon^2}{4} \left( \sigma^2 \xi^{- (1 + 3 \omega)} - 1 \right) }
~~~~,~~(\text{closed model})
\label{ClosedModel}
\end{eqnarray}

It is convenient to parameterize the equation of state variable as,
$\omega(n) := \Case{3 - n}{3(n - 1)}$ so that $n = 0$ corresponds to the
cosmological constant, $n = 2$ corresponds to the radiation while $n =
3$ corresponds to the dust.

For the flat models, the substitution $\xi := (\Case{\sigma \epsilon}{2}
\text{cosh} \lambda)^{n - 1}$, leads to,
\begin{eqnarray}\label{FlatParametric}
\left[\frac{\sigma}{n - 1}
\left(\frac{2}{\sigma \epsilon}\right)^n \right](\tau - \tau_0) & = & 
\int (\text{cosh} \lambda)^n d \lambda \nonumber \\
& = & 2^{-n} \sum_{r = 0}^n \ ^n C_r \left(\frac{\text{sinh}(2 r - n)\lambda}{2
r - n}\right) %\\
\end{eqnarray}

For even integer $n$, the parenthesis in the $r = n/2$ term in the
summation, is to be replaced by $\lambda$.  We have thus obtained the
solution of the Friedmann equation (\ref{FlatModel}) in a parametric
form, in particular for the cases of interest namely $n = 0, 2, 3$. 

To recover the $\epsilon \to 0$ solutions, one has only to note that for
{\em non-zero} $n$, the left hand side of (\ref{FlatParametric})
diverges for $\tau > \tau_0$ (say), and implies that one must have
$\lambda$ very large. The leading term on the right hand side is then
$\Case{2^{-n}}{n} e^{n \lambda}$. Likewise, for large $\lambda$, $\xi
\sim (\Case{\sigma \epsilon e^{\lambda}}{4})^{n - 1}$. One can now
eliminate $\epsilon e^{\lambda}$ to get $a(t) \propto (t - t_0)^{\Case{n
- 1}{n}}$. Clearly, these match with the usual ($\epsilon = 0$)
solutions. The case of the cosmological constant ($n = 0$) is discussed
in the next subsection.

For the closed models, again one can obtain parametric form of the
solutions which involves an integration. However the qualitative behavior
can be seen easily. Now a different substitution is convenient. 
Putting $\xi := (\Case{\sigma}{\text{cosh} \lambda})^{(n - 1)}$ one obtains,
\begin{eqnarray} \label{ClosedParametric}
- \frac{1}{n - 1} \sigma^{1 - n} (\tau(\lambda) - \tau_0) & = &
\int_{\lambda_0}^{\lambda} d \lambda'
\frac{(\text{cosh}\lambda')^{-n}}{\sqrt{1 - \left(\frac{\epsilon^2}{4}\right)
(\text{sinh} \lambda'})^2}
\end{eqnarray}

Clearly, $\lambda$ must be bounded to ensure $0 \le \text{sinh}^2
\lambda \le \Case{4}{\epsilon^2}$. One can also check that the integrand
is integrable at the maximum value of $\lambda$. Taking $\tau_0 = 0 =
\lambda_0$ and $\tau := \tau_{\text{max}}$ for the upper limit of
integration to be $\lambda_{\text{max}}$ (defined by
$\text{sinh}^2\lambda_{\text{max}} = \Case{4}{\epsilon^2}$), one obtains
an oscillatory universe with a finite period given by $T := 2
\tau_{\text{max}}$ for all the three cases of interest. The 
$\epsilon \to 0$ limit is simpler in this case since $\xi$ does not
depend explicitly on $\epsilon$. Also $\lambda_{\text{max}}$ diverges
and therefore for any finite $\lambda$ we can simply take $\epsilon = 0$
in the integrand. For $n = 2, 3$, then one recovers the usual
(parametric) form of the solutions. 

For dust and radiation, the corrections due to non-zero $\epsilon$ are
extremely small. This is to be expected since the densities decrease
with the scale factor, making the $a^{-1} H_m^2$ term very small.

\subsection{Cosmological Constant}
\label{CosmoConst}
This is obtained by taking $H_m := \Lambda p^{3/2} = \Lambda
a^3/8$ and as mentioned before, mathematically it corresponds to $n =
0$. In the equation (\ref{DimlessDefns}), we replace $\bar{\rho}$ by
$\Lambda$. 

For $\epsilon = 0$, one has the well known solutions: $a(t) =
a(0)\ {\mathrm{exp}}(\sqrt{\Lambda/6}\ t)$ for $\eta = 0$ and $a(t) =
\sqrt{6/\Lambda}\ {\mathrm{cosh}}(\sqrt{\Lambda/6}\ (t - t_0) \ )$ 
for $\eta = 1$.

For $\epsilon \ne 0$, the solutions are qualitatively different. For the
spatially flat case, $\eta = 0$, one has a re-collapsing solution given
explicitly as, 
\begin{equation}\label{FlatSoln} 
a(t) ~=~ \frac{a_{{\mathrm{max}}}}
{{\mathrm{cosh}}(\sqrt{\Lambda/6}\ (t_0 - t)\ )} ~~~,~~~ a^2_{{\mathrm{max}}} :=
\frac{24}{\epsilon^2 \Lambda} \ .
\end{equation}  

The scale factor vanishes only for $t \to \pm \infty$ and the solution 
is non-singular. As $\epsilon \to 0$, the maximum value for scale factor 
diverges and one is either in the expanding or contracting phase. One
can obtain the $\epsilon = 0$ solution by taking the limit $\epsilon \
\to 0 ,\  t_0 \to \infty$ with $\epsilon e^{\sqrt{\Case{\Lambda}{6}}t_0}$
held constant.

For the closed model, $\eta = 1$, the scale factor is bounded from both above
and below as $1 \le \Lambda a^2/6 \le (1 + 4/\epsilon^2)$ and the
universe keeps oscillating between these values in {\em finite} time,
with the period of oscillation given by,
\begin{equation}\label{ClosedPeriod} T ~=~ 2 \sqrt{6/\Lambda}
\int_0^{\lambda_{{\mathrm{max}}}} \frac{d \lambda}{ \sqrt{1 -
\frac{\epsilon^2}{4} {\mathrm{sinh}}^2 \lambda} } ~~~,~~~
{\mathrm{sinh}}(\lambda_{{\mathrm{max}}}) := 2/\epsilon \ .
\end{equation}

As $\epsilon \to 0$, the maximum value of $\xi$ diverges, the period
diverges. The usual de Sitter solution is recovered by just taking
$\epsilon = 0$ in the integrand of equation (\ref{ClosedParametric})
which gives $\lambda \sim \sigma (\tau - \tau_0)$ and eliminating
$\lambda$.  In both the cases above, one can check from the Raychoudhuri
equation that smaller scale factor is indeed a local minimum while the
larger one is a local maximum (as a function of $t$).

For the cosmological constant as the only source, for $\epsilon = 0$, 
the equation of state, $w := P/\rho = -1$ remains independent of the
scale factor and the Raychoudhuri equation then implies an
always accelerating universe, precluding the possibility of re-collapse.
With the $\epsilon$ corrections incorporated, the effective density and
pressure is modified and the modified equation of state acquires a scale
factor dependence which is such that there is always a decelerating
phase before re-collapse.

Had we taken $\epsilon \to 0$, we would have got the classical GR
Hamiltonian leading to the usual exponentially expanding universe
without a re-collapse ($\eta = 0$).  The difference equation on the
other hand shows that discrete wave function, in any {\em one sector} begins
to oscillate over the quantum geometry scale $\sim \Case{4}{6}\mu_0
\gamma \lP^2$ for large values of the triad, implying a breakdown of
pre-classicality (no small scale variations) of the exact solutions.
While deviations from pre-classicality are expected in the small volume
regime due to quantum effects, deviations in the very large volume
regime have no such physical reasons and are viewed as an `infra-red
problem' \cite{IsoCosmo,Infrared}. A typical example of such large
volume deviations is that of cosmological constant (and non-zero spatial
curvature). No quantity which has a local interpretation such as energy
density ($\Lambda$) is ill-behaved and the integrated quantity diverges
due to integration over infinite volume. In such cases, the homogeneous
idealization (integrand being a constant) is thought to become
in-applicable beyond a finite volume. Noting that the wave function has
an oscillation length of $(\Lambda a)^{-1}$ \cite{IsoCosmo}, one can
obtain a bound on the largest scale factor, $\tilde{a}$, by requiring
the oscillation length to be larger than the scale of slow variation,
$\sqrt{q} = \sqrt{\Case{\epsilon}{3}} \lP$. This gives, $\tilde{a}
\lesssim \sqrt{\Case{3}{\epsilon}} (\Lambda \lP)^{-1}$.

In the effective picture, we have a maximum value, $a_{\text{max}}$ of
the scale factor due to re-collapse and for $\eta = 0$ it is given by
$a_{\text{max}}^2 = \Case{24}{\epsilon^2 \Lambda}$ in equation
(\ref{FlatSoln}).  Demanding that $a_{\text{max}} \le \tilde{a}$ gives a
bound on the cosmological constant: $\Lambda \lP^2 < \epsilon/8$. For
such a cosmological constant, the effective picture can be consistent
with pre-classical behaviour of exact solutions. Incidentally, for the
currently favored value of the cosmological constant, one has $\Lambda
\lP^2 \sim 10^{-120}$.

We would also like to note that in \cite{NPV} re-collapse of a
$\Lambda-$dominated universe has been obtained from a different
perspective. These authors note that the discrete wave functions
cease to show oscillatory behaviour beyond a critical volume.  The value
of the critical volume matches exactly with the $a_{\text{max}}$ above
thus corroborating the WKB intuition \footnote{GD would like to thank
Alejandro Perez and Kavin Vandersloot for e-mail exchanges regarding
\cite{NPV}.}. 

In the above discussion, we have used the large scale factor expressions
for the equations, mainly because we were interested in seeing the
re-collapse possibility. Secondly we also used the effective classical
picture right close to the WKB turning point. For inferring qualitative
implications as done above, this is okay. The full effective classical
equations together with the domain of validity of the WKB approximation
have been given in the previous section. The main lesson is that the
corrections due to discreteness can change some of the solutions
qualitatively while for others these are perturbative in nature.

\section{Discussion and Summary}
\label{DiscussionSection}

Let us recapitulate the key steps in the derivation of the effective
Hamiltonian and clarify the role of WKB approximation and its
interpretation.

We began from the quantum Hamiltonian constraint (\ref{FunctionalEqn}).
While it has a variety of solutions, we assumed that there exist a class
of solutions which are Taylor expandable, at least over some bounded
interval(s) of the triad eigenvalues. This allowed us to introduce the
$\delta C_{\pm}, \delta\Phi_{\pm}$ in equations (\ref{TaylorExpanssion},
\ref{DeltaAmplitude}, \ref{DeltaPhase}). Next, we made the further
assumption of slowly varying amplitude and phase to {\em truncate} the
$\delta C_{\pm}, \delta\Phi_{\pm}$ to first derivative of the amplitude
and second derivative of the phase. Now the WKB approximation is made to
arrange the expansions of the trigonometric functions according to
powers of $\hbar$ to arrive at a Hamilton-Jacobi equation
(\ref{RealEqn}) for the phase, from the ($o(\hbar^0)$) terms as well as
an equation for the amplitude (\ref{ImEqn}) from the ($o(\hbar)$) terms.
The self-consistency of the slow variation of the amplitude and the
phase can be used to infer the conditions for validity of WKB
approximation.  These were further simplified to equations
(\ref{CorrectHamJac}, \ref{WKBCondition}) in the region of interest ($p
\gg q$).  The effective Hamiltonian is read-off from the simplified
Hamilton-Jacobi equation (\ref{CorrectHamJac}). We also note that the
usual, second order Wheeler-DeWitt differential equation has been
by-passed. 

Since the higher derivatives of the amplitude and phase are also higher
orders in $\hbar$, the WKB approximation (keeping leading powers of
$\hbar$) is correlated with the assumption of slow variation. As is well
known, the WKB approximation fails near the `turning points' at which
the first derivative of the phase vanishes. For definiteness, let us
assume that these occur at some $p_{\text{min}} < p_{\text{max}}$. The
domain of validity of the effective Hamiltonian is necessarily within
these values. Since the WKB solutions themselves are only approximate
solutions of the fundamental difference equation, it is possible that
these solutions fail to be a good approximation to the exact solutions.
This can happen, for example, when {\em all} the exact solutions fail to
be slowly varying (while WKB solutions are slowly varying). Let us
assume that some particular exact solution is well approximated by the
WKB solution in an interval $(p'_{\text{min}}, p'_{\text{max}})$. The
domain of relevance of the effective Hamiltonian is then the
intersection of these two intervals. The intersection must be non-empty
since there do exist exact solutions which are slowly varying
exhibiting classical behavior. Clearly, the largest possible domain
for effective Hamiltonian is the interval between the two turning
points: $p_{\text{min}}, p_{\text{max}}$. Since, as explained in the
section (\ref{LQCSummarySection}),  exact solutions are countable sums
of independent solutions of the difference equation, it should be
possible to choose linear combinations which approximate any given WKB
solution over the WKB domain, for example as a least square fit. Precise
and useful quantification of a WKB solution approximating an exact
solution however needs a separate work.

To construct WKB solutions, one appeals to the connection of
Hamilton-Jacobi equation with an associated (classical) Hamiltonian
dynamics. It is well known (in a general context) that any given
solution, $\Phi(q^i)$, of a Hamilton-Jacobi equation selects a
sub-class of Hamiltonian trajectories in the phase space, namely, those
with initial conditions $(q^i(0), p_i(0) := \Case{\partial\Phi}{\partial
q^i}|_0)$. Conversely, a {\em local} solution, in a neighbourhood of a
point ($q^i_0$), is constructed as $\Phi(\ q^i(t, q^i_0, \dot{q}^i_0)\ )
:= \int_0^t L(\ q^i(t'), \dot{q}^i(t')\ ) dt'$. Here, $q^i(t, q^i_0,
\dot{q}^i_0)$ denotes a family of solutions of the Euler-Lagrange
equation (obtained from the solutions of the Hamilton's equations of
motion), with initial conditions $q^i_0, \dot{q}^i_0$. Thus, the
solutions of the effective Hamiltonian equations of motion can be used
to construct corresponding (local) solutions of the Hamilton-Jacobi
equation and hence the amplitude and the phase. In this manner, the
effective Hamiltonian can be used to construct solutions which are to be
approximated by exact solutions of the fundamental equation
(\ref{DifferenceEqn}). 

In summary, at least in the context of isotropic LQC, the discreteness
corrections are manifested at the level of the effective Hamiltonian.
These corrections arise due to a discrete geometry underlying the
continuous one and are controlled by the parameter $\mu_0$ and
the Barbero-Immirzi parameter $\gamma$. For the cases where $a^{-1}H_m$
grows with the scale factor, the corrections alter the qualitative
behaviour while in other cases the corrections are perturbative in
nature. 

\begin{acknowledgments} 

We thank Martin Bojowald, Golam Hossain, Jorge Zanelli for helpful,
critical comments and Abhay Ashtekar, Parampreet Singh, Kevin
Vandersloot and Alok Laddha for discussions. GD would like to thank
Martin Bojowald for an invitation to the Max Planck Institut f\"ur
Gravitationsphysics, Golm and Abhay Ashtekar for an invitation to the
Institute for Gravitational Physics and Geometry, State College. Part of
this work was done during these visits and the warm hospitality is
gratefully acknowledged.

\end{acknowledgments}

%%%%%%%%%%%%%%%%%%%%%%%%%%%%%%%%%%%%%%%%%%%%%%%%%%%%%%%%%%%%%%%%%%%%%%%

\end{document}